\documentclass{WileyMSP-template}
\usepackage{amsmath,amssymb}
\usepackage{hyperref}
\hypersetup{
    colorlinks = true,
    linkcolor =blue,
    urlcolor = blue
}

\begin{document}

\pagestyle{fancy}
\rhead{\includegraphics[width=2.5cm]{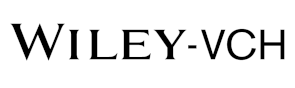}}

\title{
A first application of machine and deep learning for background rejection in the ALPS II TES detector }

\maketitle

\author{Manuel Meyer*}
\author{Katharina Isleif}
\author{Friederike Januschek}
\author{Axel Lindner}
\author{Gulden Othman}
\author{Jos\'e Alejandro Rubiera Gimeno}
\author{Christina Schwemmbauer}
\author{Matthias Schott}
\author{Rikhav Shah}
\author{for the ALPS Collaboration}


\dedication{}

\begin{affiliations}
Dr. M. Meyer\\
Institut f\"ur Experimentalphysik, Universit\"at Hamburg, Luruper Chaussee 149, 22761, Hamburg, Germany; now at CP3-Origins, University of Southern Denmark, Campusvej 55, 5230 Odense M, Denmark\\
Email Address: mey@sdu.dk

G. Othman, PhD\\
Institut f\"ur Experimentalphysik, Universit\"at Hamburg, Luruper Chaussee 149, 22761, Hamburg, Germany\\

Dr. K. Isleif\\
Deutsches Elektronen-Synchrotron DESY, Notkestr. 85, 22607 Hamburg, Germany\\Now at Helmut-Schmidt-University, Holstenhofweg 85, 22043 Hamburg

Dr. F. Januschek, Dr. A. Lindner, J. A. Rubiera Gimeno, C. Schwemmbauer\\
Deutsches Elektronen-Synchrotron DESY, Notkestr. 85, 22607 Hamburg, Germany

Prof. M. Schott, Dr. R. Shah\\
Johannes Gutenberg-Universit\"at Mainz, Staudingerweg 7, 55128 Mainz, Germany

\end{affiliations}


\keywords{Transition Edge Sensor, Cryogenic single-photon detection, 
Pulse characterization, Backgrounds, Machine Learning, Axions}

\begin{abstract}

Axions and axion-like particles are hypothetical particles predicted in extensions of the standard model and are promising cold dark matter candidates. The Any Light Particle Search (ALPS II) experiment is a light-shining-through-the-wall experiment that aims to produce these particles from a strong light source and magnetic field and subsequently detect them through a reconversion into photons. With an expected rate $\sim$ 1 photon per day, a sensitive detection scheme needs to be employed and characterized. One foreseen detector is based on a transition edge sensor (TES). Here, we investigate machine and deep learning algorithms for the rejection of background events recorded with the TES. We also present a first application of convolutional neural networks to classify time series data measured with the TES.

\end{abstract}


\section{Introduction}

Axions and axion-like particles (ALPs) are hypothetical particles predicted in extensions of the Standard Model of particle physics~\cite{2018PrPNP.102...89I}.
Both axions and ALPs are candidates to explain the observed density of cold dark matter in the Universe~\cite{1983PhLB..120..133A,1983PhLB..120..137D,2012JCAP...06..013A}. Additionally, axions could  solve the so-called strong CP~problem of the strong interactions~\cite{1977PhRvD..16.1791P,1978PhRvL..40..223W,1978PhRvL..40..279W}.
One predicted interaction of axions and ALPs is the conversion into photons in the presence of external magnetic fields.
Such an interaction would make it possible to detect axions and ALPs present in the dark matter halo in the Milky Way or produced in astrophysical sources such as the Sun or in supernova explosions~\cite{2018PrPNP.102...89I}.

In contrast to searches relying on astrophysical sources of ALPs, the Any Light Particle Search II (ALPS~II) experiment aims to produce and subsequently detect ALPs with the so-called light-shining-though-a-wall (LSW) technique~\cite{1983PhRvL..51.1415S,2013JInst...8.9001B,2022MUPB...77..120I}.
In ALPS~II, a powerful laser beam 
is immersed in a strong magnetic field and directed onto an opaque barrier.
A fraction of photons in the laser beam convert to ALPs, which traverse the barrier unimpeded. 
Behind this wall, in an additional magnetic field, ALPs reconvert into photons with the same properties as the original ones, which can be subsequently detected. 
Once commissioned, ALPS~II will reach unprecedented sensitivity for an LSW-type experiment by employing a high-power infrared laser at a wavelength of 1064\,nm, optical cavities for additional power build-up before and behind the wall, and sensitive photon detectors measuring rates down to $\sim10^{-6}\,$Hz~\cite{2022epsc.confE.801S,Hallal:2020ibe}. 
Within a 20\,day measurement we aim to to probe 
photon-ALP couplings down to $g_{a\gamma} \gtrsim 2\times10^{-11}\,\mathrm{GeV}^{-1}$ for masses $m_a \lesssim 10^{-4}$\,eV.
This would make it possible to probe ALP dark matter scenarios \cite{2021JHEP...01..172C} and axion models that predict a large coupling to photons \cite{2017JHEP...01..095F,2021JHEP...06..123S}.
For this photon-ALP coupling, we expect a reconverted photon rate of $n_s \gtrsim\times10^{-5}\,$Hz (corresponding $\sim 1\,$photons per day) given the ALPS~II design specifications.
To significantly detect such a low rate, the background rate has to be $\lesssim 10^{-5}\,$Hz~\cite{2022epsc.confE.801S}.

One foreseen detection technique is based on a transition edge sensor (TES)~\cite{2023NIMPA104667588R}. 
Such sensors are essentially microcalorimeters: 
they consist of a superconducting chip integrated in a circuit where they are biased at a temperature between the normal and superconducting phase~\cite{2005cpd..book...63I}.
A reconverted photon will be guided via an optical fiber to the TES where it is absorbed.
This increases the chip's temperature thereby causing a large change of its resistance of the order of several Ohms. 
Through an inductive coil, the current change induced by the change in resistance leads to a change in the magnetic field, which is read out with a superconducting quantum interference device (SQUID). 
Such detectors can be optimized for near-infrared light and show high quantum efficiencies close to unity, a high energy resolution, and low dead time~\cite{2008OExpr..16.3032L,2022JLTP..tmp...93S}.

The majority of background events registered with the TES is expected from thermal radiation of the warm (at room temperature) end of the optical fiber~\cite{miller2007}. We call this background source \textit{extrinsic}.
Additional sources of background include radioactive decays inside the detector volume and energy deposition of charged cosmic rays interacting with the TES or the surrounding material~(e.g., Refs.~\cite{lotti2012,lotti2014}).
We refer to these types of events, which are present with and without an optical fiber, as \textit{intrinsic} background events. 
To achieve the necessary low background rates, background events must be efficiently rejected by both the experimental design (see, e.g., Ref.~\cite{lopez2015}) and the data analysis. 

Here, we present a first investigation of the performance of machine learning (ML) and deep learning (DL) classification algorithms to discriminate fake signals from intrinsic background events at the data analysis level.
Due to the excellent performance in, e.g., classification tasks, both ML and DL algorithms enjoy increasing popularity in fundamental physics research as a whole~\cite{2022NatRP...4..399K}
and for searches of axion signatures in particular~\cite{2020JCAP...03..046D,2021JHEP...11..138R,2022PDU....3701118K}.
As we will see in Section~\ref{sec:data}, where we introduce the training data for our classifiers, the TES data are essentially time series in which individual photons are seen as pulses.
The integral over this pulse is proportional to the deposited energy and thus the photon energy~\cite{2005cpd..book...63I}. 
Therefore, the signal-and-background discrimination  boils down to a time series classification (TSC).
In particular DL algorithms perform particularly well for such tasks~\cite{IsmailFawaz2019}.
In previous analyses of ALPS~II TES calibration data, signal and background events were distinguished through a standard pulse shape analysis (PSA)~\cite{2015JMOp...62.1132D,2022epsc.confE.801S,2022JLTP..tmp...93S}.
In PSA, recorded pulses are fit either with a parametric function or a template pulse with a free amplitude parameter. 
The distinction between signal and background is then achieved through cuts in the parameter space of the extracted pulse parameters, i.e., extracted \textit{features} (e.g., pulse amplitude and pulse integral). 
In principle, ML and DL algorithms should be perfectly suited to either optimize such cuts or to find high-dimensional data representations where the feature space of signal and background events can be separated in an optimal way (in the sense of minimizing some cost function). 
This will be explored in Section~\ref{sec:features}.
Instead of feature extraction we will use the time lines themselves for classification in Section~\ref{sec:cnn}.
We closely follow Ref.~\cite{IsmailFawaz2019} and present first results of convolutional neural networks (CNNs) for this task.
Compared to conventional (fully-connected) deep neural networks, CNNs are based on shared weights from convolutional kernels, which reduced the number of parameters and leads to an improved learning of trans\-lation-equi\-variant features.
The results of both strategies are presented in Section~\ref{sec:results}.
In Section~\ref{sec:conclusions}, we provide conclusions and an outlook on how to improve the present proof-of-concept study and how to extend it in the future. 

\section{Data for Classifier Training}
\label{sec:data}

\begin{figure}
    \centering
    \includegraphics[width=.32\linewidth]{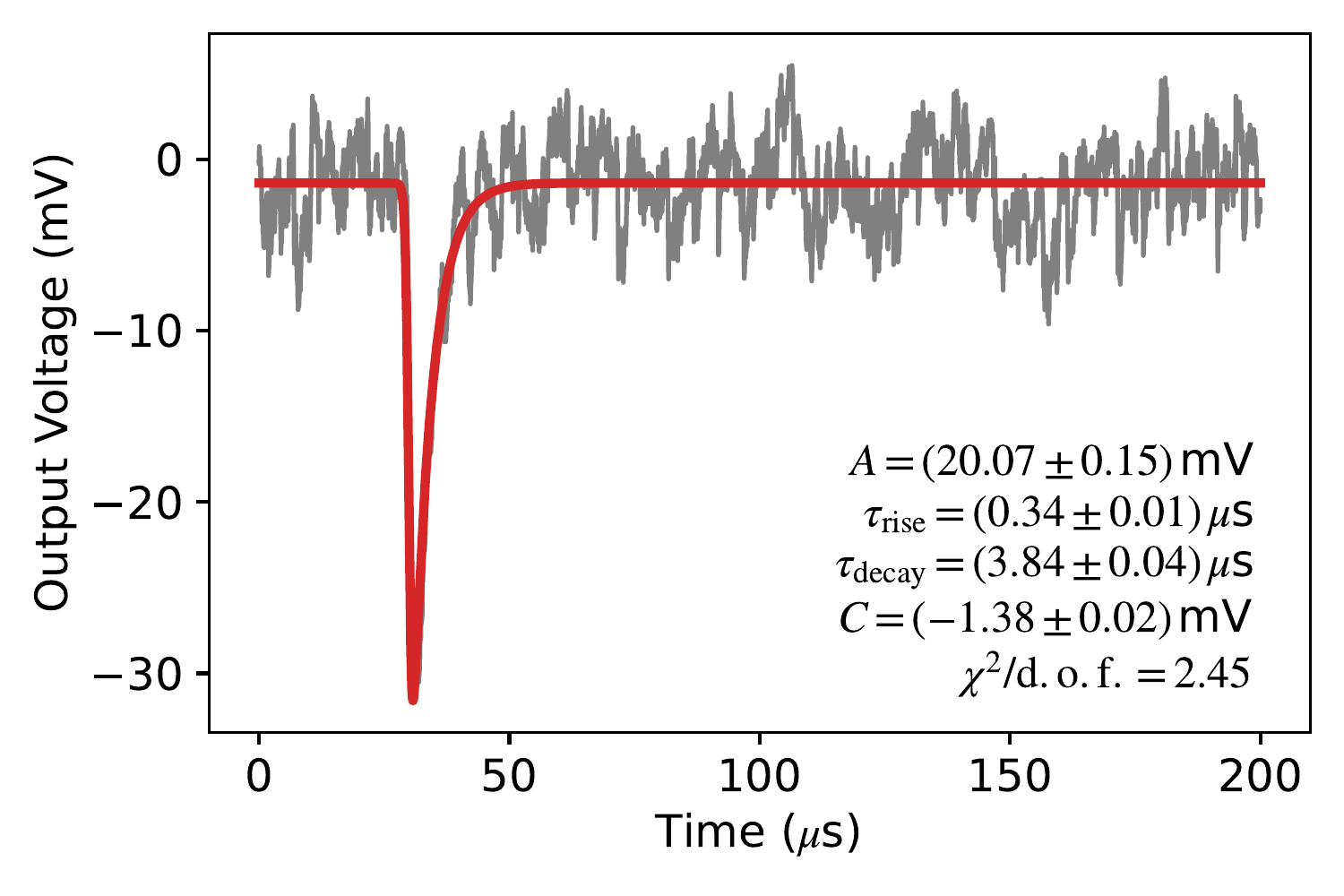}
    \includegraphics[width=.32\linewidth]{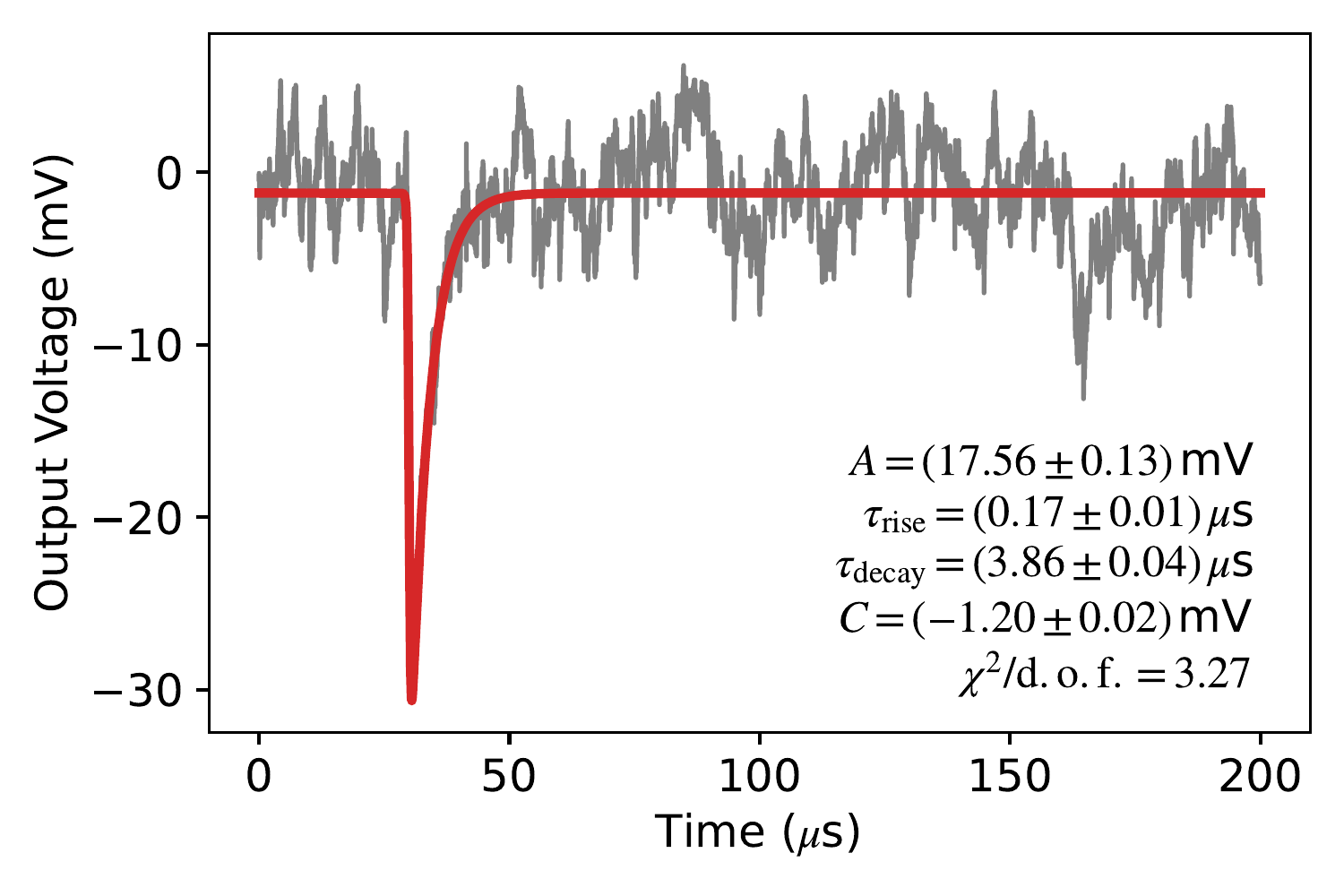}
    \includegraphics[width=.32\linewidth]{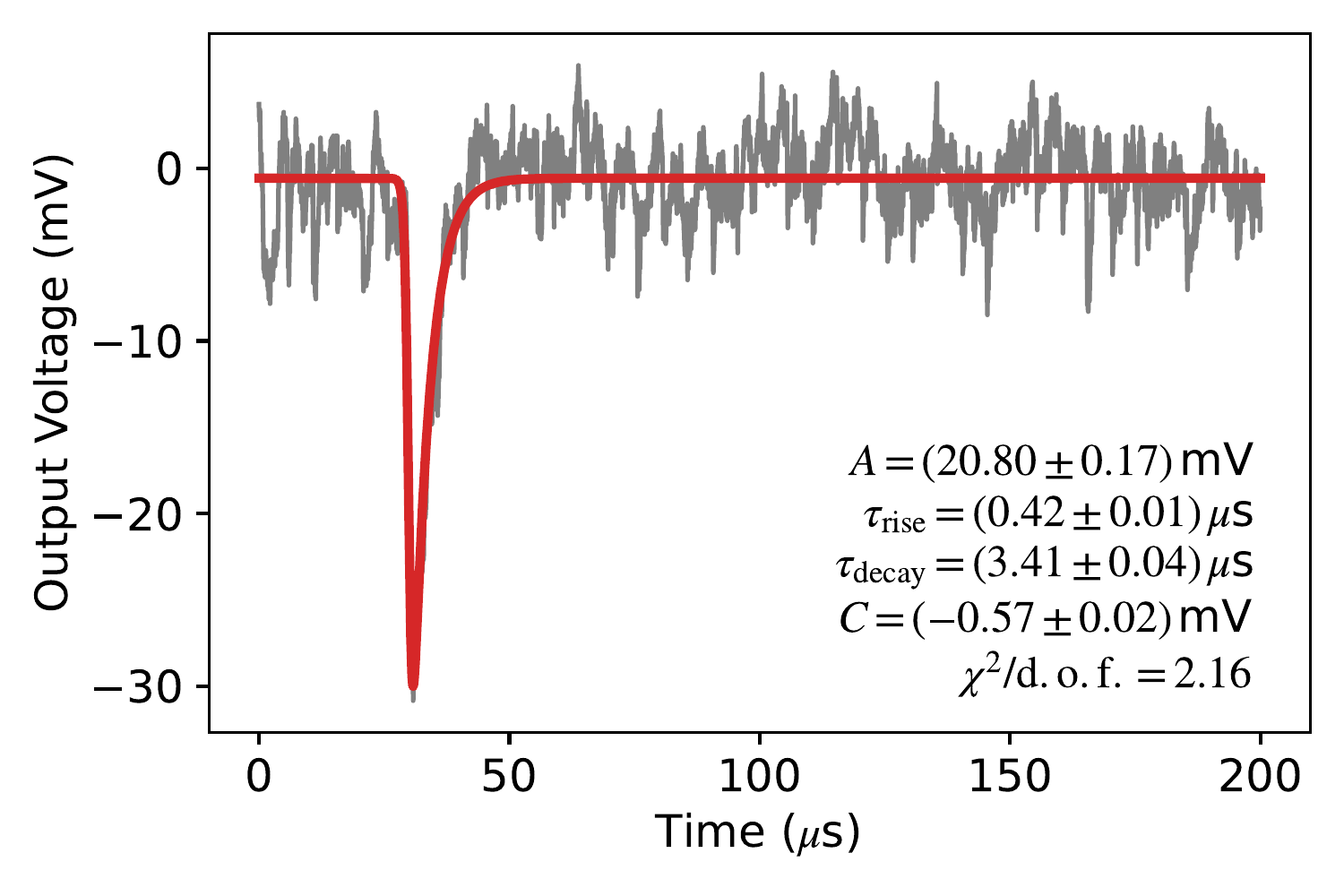}
    
    \includegraphics[width=.32\linewidth]{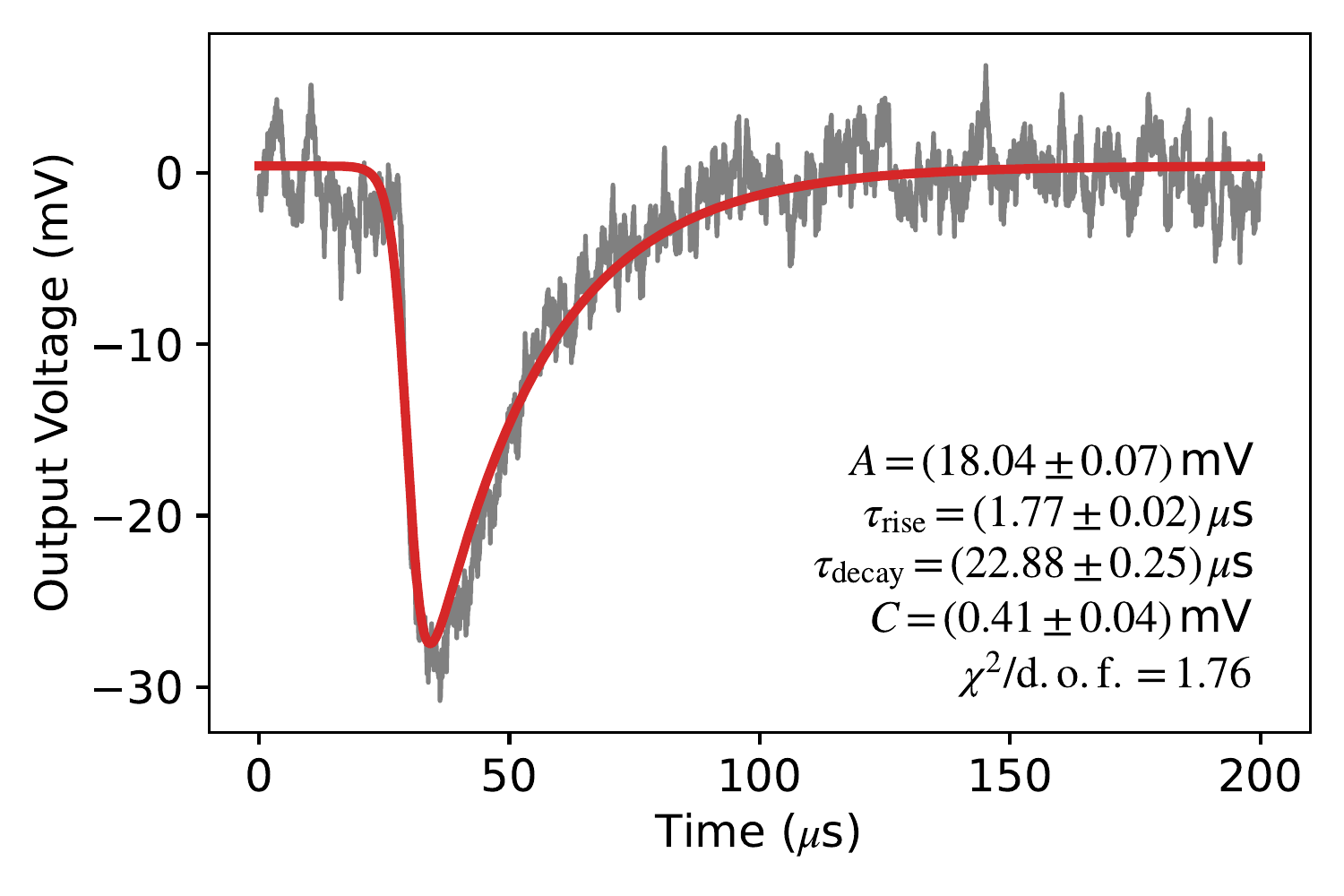}
    \includegraphics[width=.32\linewidth]{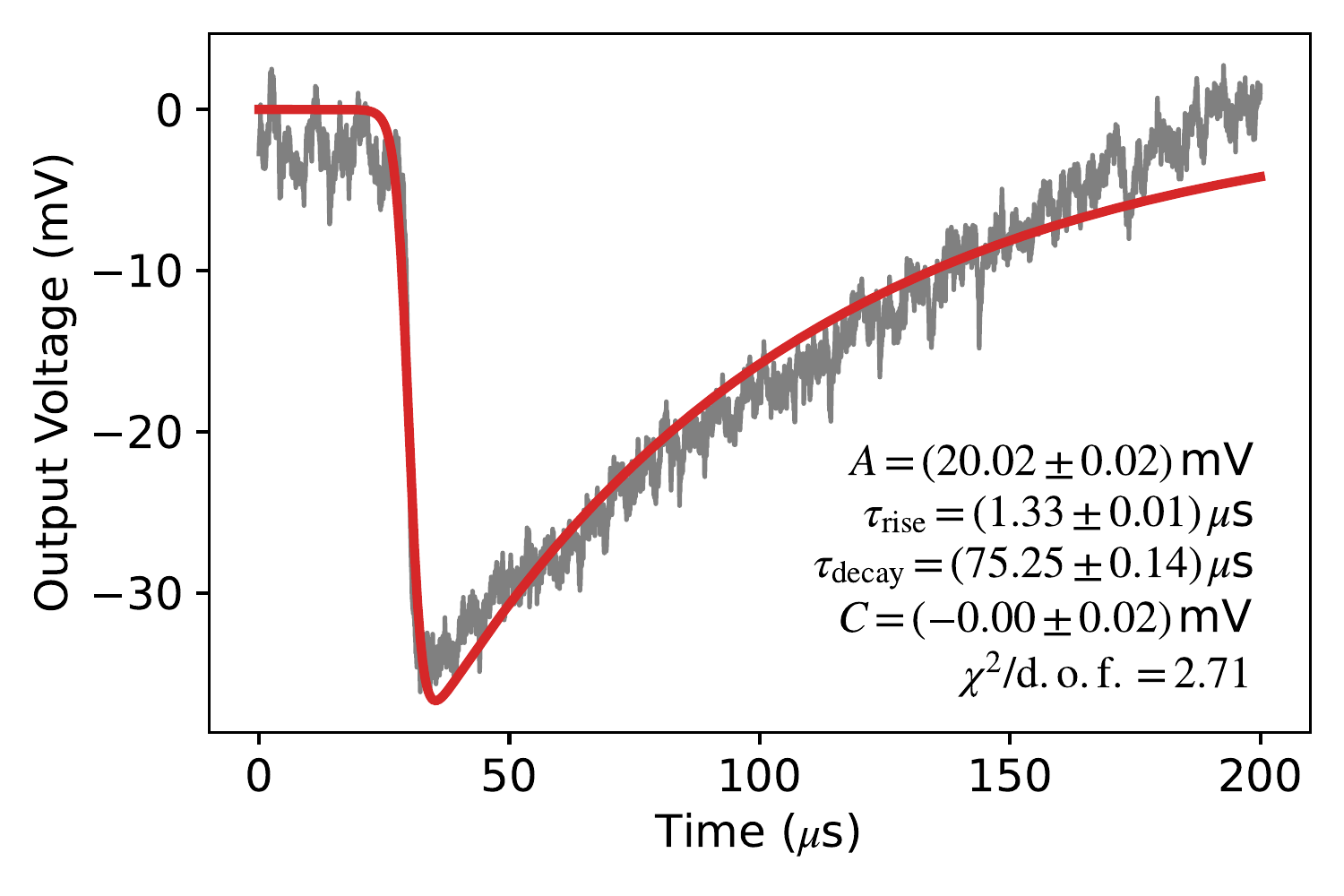}
    \includegraphics[width=.32\linewidth]{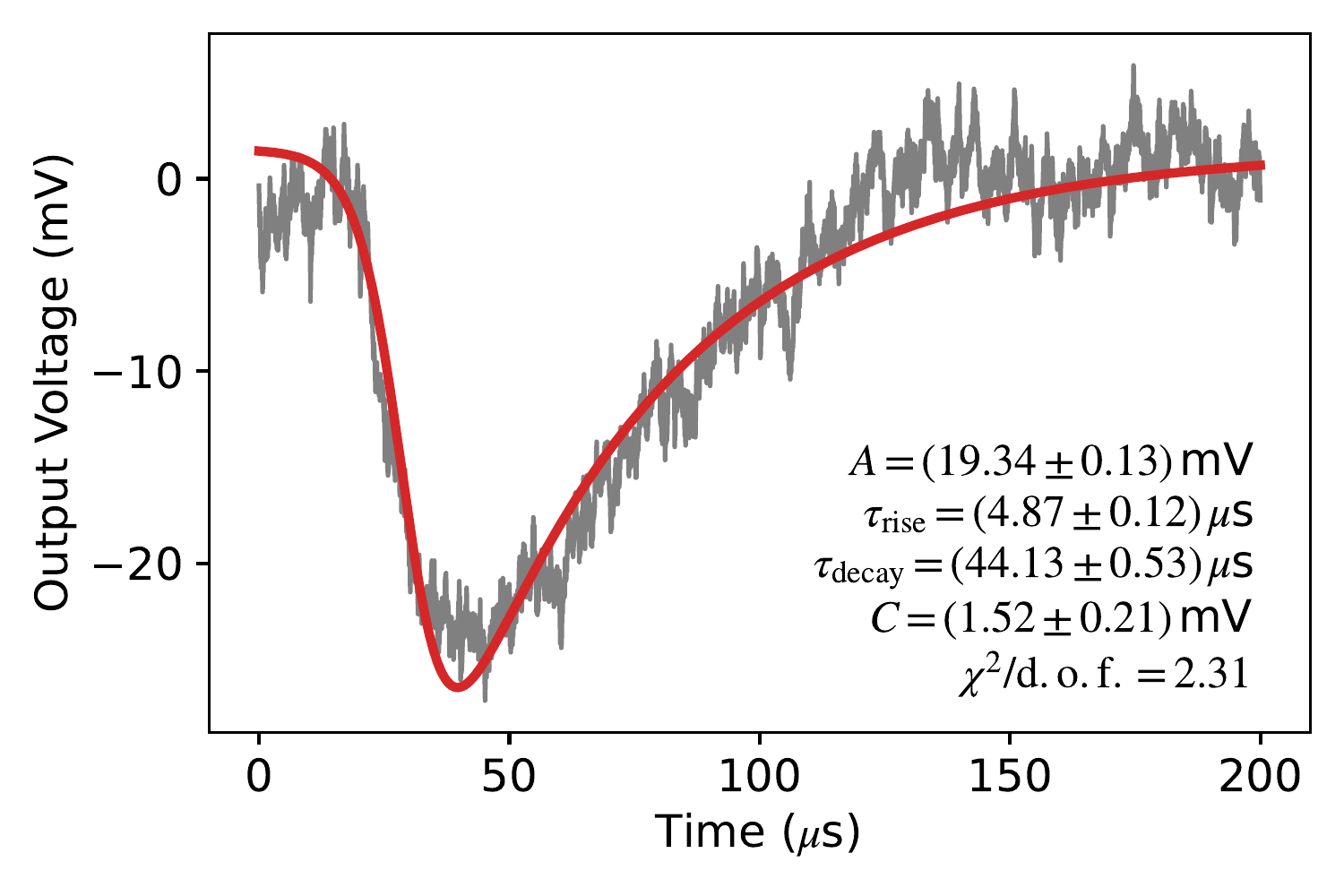}
    \caption{Example traces recorded with the TES. \textit{Upper panels:} Time lines triggered by 1064\,nm laser photons. \textit{Lower panels:} examples of intrinsic background events recorded while the optical fiber was disconnected from the TES.}
    \label{fig:ex-event}
\end{figure}

For training the classifiers, we use the same  data sets as described in Refs.~\cite{2022epsc.confE.801S, 2022JLTP..tmp...93S} which were collected in an experimental setup for characterizing the TES. 
In particular, intrinsic background events were collected in a continuous data run lasting $T = 518\,$hours, in which the TES was not connected to an optical fiber. 
These background events are labeled $y = 0$.
In a second data run, 
real photon signals were generated by connecting a continuous wave laser at a wavelength of about 1064\,nm to an optical fiber which was then attached to the TES (class labels $y = 1$).
This data run lasted for less than a minute given the high photon rate of the input laser. 
Each event $i$ consists of a voltage time line (sometimes called trace) with $M$ sample points, measured with the TES and SQUID setup, $x_i \equiv (x_{i1}, \ldots, x_{iM})^T$.
Events were triggered and saved to disk when the amplitude reached a trigger threshold $< -20\,$mV.
This threshold is chosen as a compromise between the reduction of background events while loosing close to zero events due to 1064\,nm photons.
Each trigger window is 200$\,\mu$s long (including 30\,$\mu$s before the trigger time) with a sampling rate of $f_\mathrm{sample} = 50\,$MHz yielding $M = 10^4$ samples per trace.
We show example traces triggered by a laser photon in the upper panels of Fig.~\ref{fig:ex-event} and traces from intrinsic background events in the lower panels of Fig.~\ref{fig:ex-event}.
For the chosen examples, it is easy to distinguish light from background events by eye when comparing the overall pulse shapes. 

The time lines are fit with an exponential rise and decay function $V(t)$ \footnote{We prefer this phenomenological function over the pulse shape from small signal theory~\cite{2005cpd..book...63I} as it is continuous for all values of $t$. It is commonly used to described the time variability of certain galaxies, see e.g., Ref.~\cite{2019ApJ...877...39M}.}
\begin{equation}
    V(t) = C - 2 A \left[\exp\left(\frac{t_0 - t}{\tau_\mathrm{rise}}\right) + \exp\left(\frac{t - t_0}{\tau_\mathrm{decay}}\right)\right]^{-1},\label{eq:pulse}
\end{equation}
using a $\chi^2$ minimization. The parameters of the function are the pulse normalization $A$, the trigger time $t_0$, the rise and decay times $\tau_\mathrm{rise,decay}$ , respectively, and a constant offset $C$. 
The rise and decay times are connected to the electrical and thermal constants of the TES circuit~\cite{2005cpd..book...63I}.  
For $t = t_0$, One finds that $V(t_0) = C - A$.
It should be noted that the pulse minimum is not reached at $t_0$ but at a later time $t_\mathrm{peak}$, where $V(t_\mathrm{peak}) = C - 2A\tau_\mathrm{rise} (\tau_\mathrm{rise} + \tau_\mathrm{decay})^{-1}(\tau_\mathrm{decay} / \tau_\mathrm{rise})^{\tau_\mathrm{decay}/(\tau_\mathrm{rise} + \tau_\mathrm{decay})}$.
For the $\chi^2$ minimization, a constant uncertainty of 1.5\,mV is assumed for each measured voltage value. 
This choice is simply motivated to achieve fast convergence of the fit. 
However, When the uncertainty is estimated from the square root of the diagonal terms of the covariance matrix of pure noise traces, similar values are found. 
Examples for the fit are also shown in Fig.~\ref{fig:ex-event} as red lines together with the best-fit values. 
After an initial minimal data cleaning of the light data,\footnote{The light data could be contaminated by background data; for this reason we exclude pulses with a decay time $\tau_\mathrm{decay} > 10\,\mu$s and a $\chi^2 / \mathrm{d.o.f.} > 6$, where d.o.f. denotes the degrees of freedom of the fit. These values are motivated from the average pulse observed in the light data.} 
we are left with in total $N=40,646$ events of which $N_\mathrm{bkg} = 39,580$ are background events recorded when the laser was off and disconnected from the TES.
For the classification based on these extracted features (Section~\ref{sec:features}), we use the best-fit values of the model in Eq.~\eqref{eq:pulse} together with the $\chi^2$ value of the fit and the integral over time of the fitted model, which we denote with $\mathrm{PI}$ (for pulse integral). 
Our feature vector thus becomes $X_i = (A, \tau_\mathrm{rise}, \tau_\mathrm{decay}, C, \chi^2, \mathrm{PI} )^T_i$ with class labels $y_i$ for samples $i=1,\ldots,N$. 
In contrast, the time series classification scheme discussed in Section~\ref{sec:cnn} will take the raw traces as input such that $X_i = x_i$ with class labels $y_i$.

\section{Training of Classifiers}

With our data at hand, we now turn to the training of the classifiers. We start with the classifiers based on the extracted time-line features in Section~\ref{sec:features} before turning to the training of a CNN on the raw time series data in Section~\ref{sec:cnn}.
Throughout, we split the data into training and test data sets using a split ratio of 80\,\% and 20\,\%. 
The classifiers will be optimized on the training set and their performance is then evaluated on the test set. 
As our data set is highly imbalanced with a ratio of $\sim40:1$ of background versus light data, we employ a stratified split of training and test data.
That means that the ratio of signal and background data is roughly the same ($40:1$) for both data sets. This ensures that we will not end up with a test data set that does not contain any light samples.

\subsection{Training of Classifiers on Extracted Features}
\label{sec:features}

We test the performance of two ML and DL algorithms for signal and background discrimination: a random forest (RF) and a multilayer perceptron (MLP), i.e., a fully connected deep neural network. 
To avoid overfitting of the MLP, $\mathrm{L}2$ regularization is applied, 
which adds the sum over all weights squared (the $\mathrm{L}2$ or Euclidean norm) to the cost function 
(see, e.g., Ref.~\cite{hastie2009elements} for a review of the different methods used in this section). 

\begin{figure}
    \centering
    \includegraphics[width=.9\linewidth]{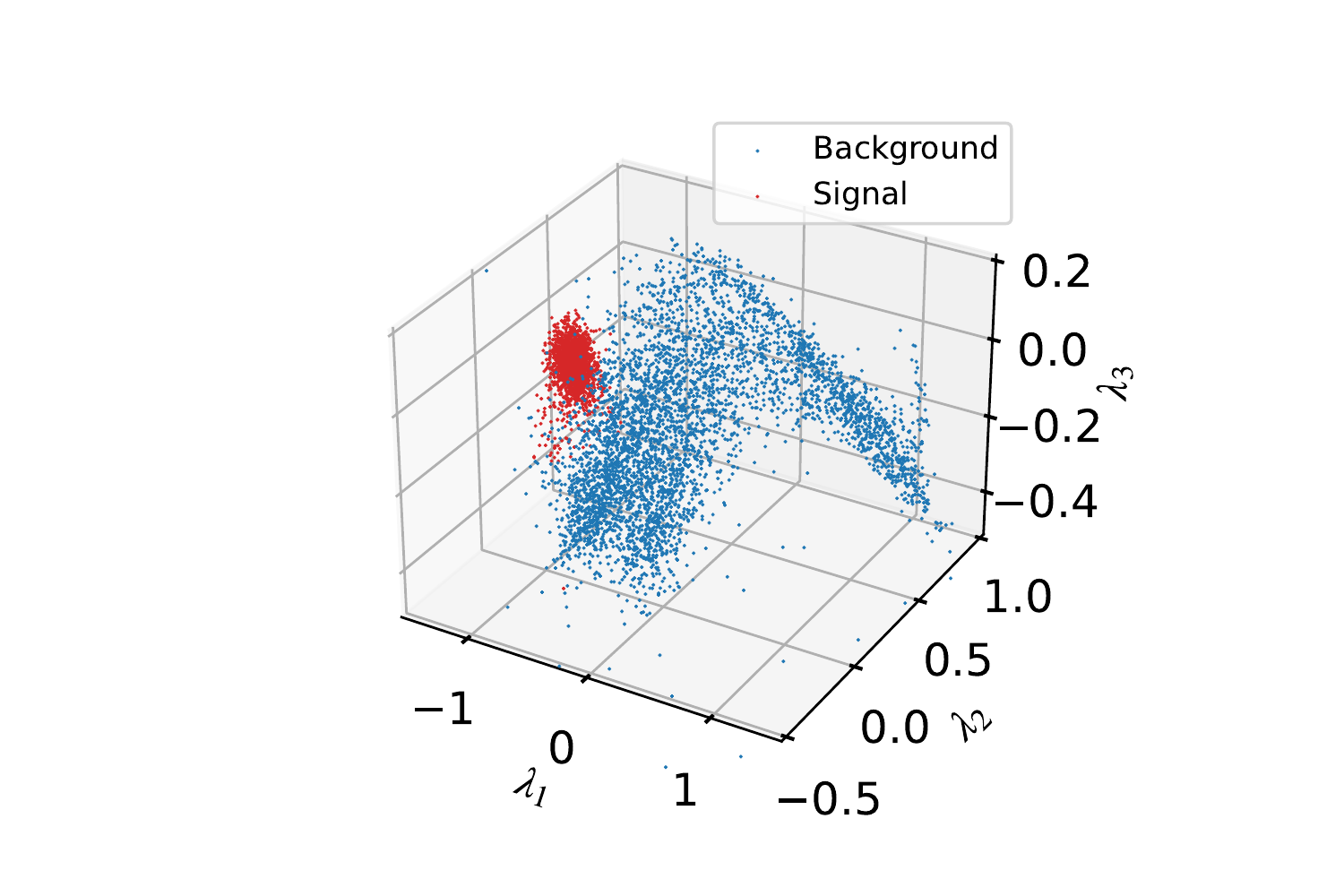}
    \caption{The first three principal components of the training data. The signal (red) and background (blue) data is already quite well separated in feature space.}
    \label{fig:pca}
\end{figure}

Before the actual training, we perform two preprocessing steps on the data. 
First, we take the logarithm of the extracted features. 
As all PI values are negative, we first multiply them with $-1$. 
Some offset values $C$ are also below zero, and we use $\log_{10}(C/(1\,\mathrm{mV}) + 1)$ for the transformation. 
Second, this log-transformed data is then further transformed using a principle component analysis (PCA)~\cite{hastie2009elements}.
The principle components are fit only to the training data and then applied to training and test data sets. 
For illustration, the first three (out of six) principle components are shown in Fig.~\ref{fig:pca}. 
The separation between signal and background events is already visible. 
We found that the log and PCA transformations resulted in better classification results and faster convergence when training the classifiers. 

Each classifier comes with its own set of hyper parameters such as the number and depth of the trees for the RF or the number of nodes and hidden layers for the MLP. 
In this first application of ML presented here, we optimize a subset of hyper parameters on coarse parameter grids to observe general trends. 
For this task we use the \texttt{scikit-learn python} package
(version 0.24.2)~\cite{scikit-learn} implementation of stratified $K$-fold cross validation~\cite{hastie2009elements} applied to the training data with $K=5$.
For the RF classifier, we change the number of trees in the  forests (100, 300, and 500 trees), the number of features to consider when looking for the best split between 1 and 6 with a step size of 1, and the minimum number of samples required to split a node between 2 and 82 with a step size of 10. 
The Gini impurity measure is used for optimizing the data splits in the trees, 
which are grown to their maximum depth.
For the MLP we consider 2, 4, and 6 hidden layers with 100 nodes per layer and values for the $\mathrm{L}2$ regularization strength $\alpha$ on a logarithmic scale between $\log_{10}(\alpha) = -4, -3.5, \ldots, -1.5$.
A rectified linear unit (ReLU) function is chosen as the activation function, and the learning rate of the MLP is held constant.
The weights of the network are found with the Adam stochastic gradient-based optimizer~\cite{kingma2014adam}.
All other hyper parameters for the RF and MLP are set to their default values in the \texttt{scikit\-learn} implementation.\footnote{
For the random forest, the minimum number of samples required to split an internal node is kept at 2 and the minimum number of samples required to be at a leaf node is kept at 1.
For the MLP, the tolerance is set to $10^{-4}$ and the learning rate is held constant at $10^{-3}$. At most, 200 epochs of learning are used.
}

The best set of hyper parameters are those that maximize
the significance $S$ of a detection of signal counts above a certain number of background events. 
For Poisson distributed data, 
the detection significance $S$ over the square root of observation time $T$ is given by~\cite{1998MPLA...13.3235B,2000NIMPA.452..518B}, 
\begin{equation}
    S / \sqrt{T} = 2\left(\sqrt{\epsilon_d\epsilon_a n_s + n_b} - \sqrt{n_b} \right).
    \label{eq:sig}
\end{equation}
In the expression above $\epsilon_d$ is the detector efficiency, $\epsilon_a$ is the analysis efficiency to correctly classify signal evens, $n_b$ is the background rate from mis-identified background events, and $n_s$ is the signal rate that depends on the photon-ALP coupling.
From the classifier predictions, $\epsilon_a$ and $n_b$ are found as follows.  
For a given threshold $\xi$, $0 \leqslant \xi \leqslant 1$, events will be classified as light-like if their predicted class label $\hat{y}_i \geqslant \xi$ (both RFs and MLPs provide predictions $\hat{y}_i$ as real numbers between 0 and 1). 
We calculate the true and false positive rates,  $\mathrm{TP}(\xi) = N_\mathrm{test}^{-1}\sum_i\left[(\hat{y}_i \geqslant \xi) \&\& (y_i == 1)\right]$ and $\mathrm{FP}(\xi) = N_\mathrm{test}^{-1}\sum_i\left[(\hat{y}_i \geqslant \xi) \&\& (y_i == 0)\right]$, respectively, 
where $N_\mathrm{test}$ is the number of samples in the test data. 
These rates are rescaled to the entire data set by multiplying with the raw trigger rate, $r_\mathrm{trig} = N_\mathrm{bkg} / T \approx 0.02\,$Hz, such that
$n_b = r_\mathrm{trig} \mathrm{FP}$. The analysis efficiency is simply equal to the true positive rate, $\epsilon_a = \mathrm{TP}$. 
For the detector efficiency, we take $\epsilon_d=0.5$ to account for potential losses in the TES sensitivity or the ALPS~II cavities and $n_s = 2.8\times10^{-5}\,$Hz. 
For choosing the best set of hyper parameters, we set $\xi = 0.5$ and compute $S$. Once the parameters are determined from $K$-fold cross validation, the classifier is re-fit on the entire training set and its score on the initial test set is evaluated. 

The whole procedure is repeated for five initial 80-20 splits of the data.\footnote{Put differently, we perform two loops. In the outer loop, we perform splits $i=1,\ldots,5$ of the whole data set into test and training sets with non-overlaping test sets. In the inner loop, a $K$-fold cross validation is performed on the training set to find the best hyper parameters, which involves another 80-20 split.} From these five splits, we calculate the median and standard deviation of $S$, $n_b$, and $\epsilon_a$ which we present in Section~\ref{sec:results}.

\subsection{A First Training of CNN on the TES Time Series Data}
\label{sec:cnn}

We also test the performance of CNNs trained on the time series data itself.
This eliminates the need for feature extraction, i.e., in our case, fitting the observed pulses with a parametric function. 
As the only preprocessing step, we perform a $z$ transformation, which is common in time series classification tasks~\cite{Bagnall2016}.
We perform the $z$ transformation on each sample individually, 
\begin{equation}
\hat{x}_i = \frac{x_i - \langle x_i \rangle}{\sqrt{\langle x_i - \langle x_i \rangle \rangle^2}},
\end{equation}
where the mean is given by $\langle x_i \rangle = M^{-1} \sum_{j=1}^M x_{ij}$.
The denominator in the expression above is the standard deviation of each time series $x_i$. 
Furthermore, to reduce memory requirements, we focus on the measurements around the trigger time  between $j=(1000,\ldots,3000)$ and downsample each time series by a factor of 4, such that $M = (3000-1000)/4 = 500$. 
Since we extract a fixed number of measurement points before and after the trigger time, it is not necessary to align the time series along the time axis as done, e.g., in Ref.~\cite{2019EPJC...79..450H}.

\begin{figure}
    \centering
    \includegraphics[width=.7\linewidth]{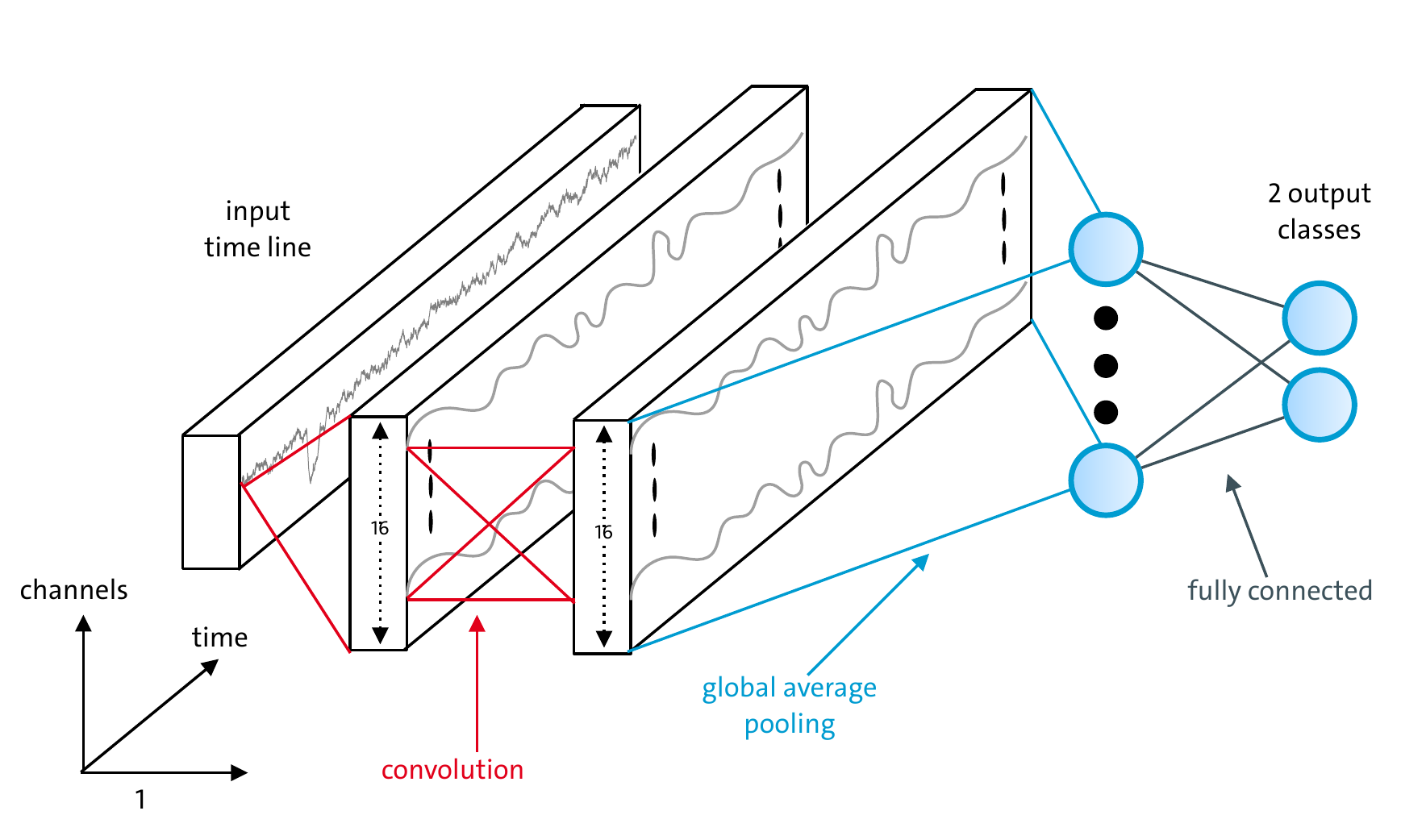}
    \caption{A sketch of our CNN architecture. Two convolutions with kernel size 11 and 16 filters are performed before a GAP layer reduces the output to 16 neurons which are connected to the 
 two output neurons (one for each class). 
 The axis labeled ``1'' denotes the direction of a forward pass within the network.
 }
    \label{fig:cnn}
\end{figure}

Our network architecture follows closely the full CNN described in Ref.~\cite{IsmailFawaz2019}. 
Specifically, we perform two convolutions with kernel size 11 with zero padding, stride equal to one, and with $N_f = 16$ filters each. 
The convolution is followed by batch normalization~\cite{ioffe2015batch} and a ReLU activation function. 
After the two convolutions, a global average pooling (GAP) is performed, which means that the time dimension is averaged over yielding 16 output neurons, one for each filter. 
The GAP output neurons are then fully connected to two output neurons--one for each class--with the categorical cross-entropy activation function.
A sketch for our simple network architecture is shown in Fig.~\ref{fig:cnn}.
The training of the network is performed with the \texttt{keras} and 
\texttt{tensorflow} packages (version 2.4.0)~\cite{chollet2015keras}.
Again, the Adam optimizer is used with an initial learning rate of 0.01.
The batch size is set to 50 and the network is trained for up to 250 epochs.
If the validation loss does not improve for 20 epochs the learning rate is reduced by a factor of 1/2 until a minimum learning rate of $10^{-4}$ is reached.
\footnote{
Given the initial learning rate, the minimum learning rate is reached after at least $\sim130$ epochs.
}
If the validation loss still does not improve after 20 additional epochs, training is stopped.
The model resulting in the minimal validation loss is saved. 

The advantage of the GAP layer is that it is possible to calculate the class activation map (CAM), which provides an easy way to visualize which portions of the time series are important for classification~\cite{Zhou2016}. In our case, the CAM itself is a univariate time series with the same dimension as the input time series.
Let $A_{f}(t)$ be the output time series after the second convolution layer (after batch normalziation and activation) for each filter $f = 1,\ldots,N_f$ and let $w_{fc}$ be the weight connecting the GAP layer node to the output class node $c = (0,1)$. Then the $\mathrm{CAM}(t)$ is given as an average over the weights, 
\begin{equation}
\mathrm{CAM}_c(t) = \sum_{f = 1}^{N_f} w_{fc} A_f(t), 
\label{eq:cam}
\end{equation}
and normalized such that $0 \leqslant \mathrm{CAM}_c(t) \leqslant 1$.
In contrast to the feature-based learning presented in Section~\ref{sec:features}, no tuning of the hyper parameters is performed, which is left for future work. 
However, the training-test split is again performed five times.

\section{Results}
\label{sec:results}

\begin{figure}[htb]
    \centering
    \includegraphics[width=.7\linewidth]{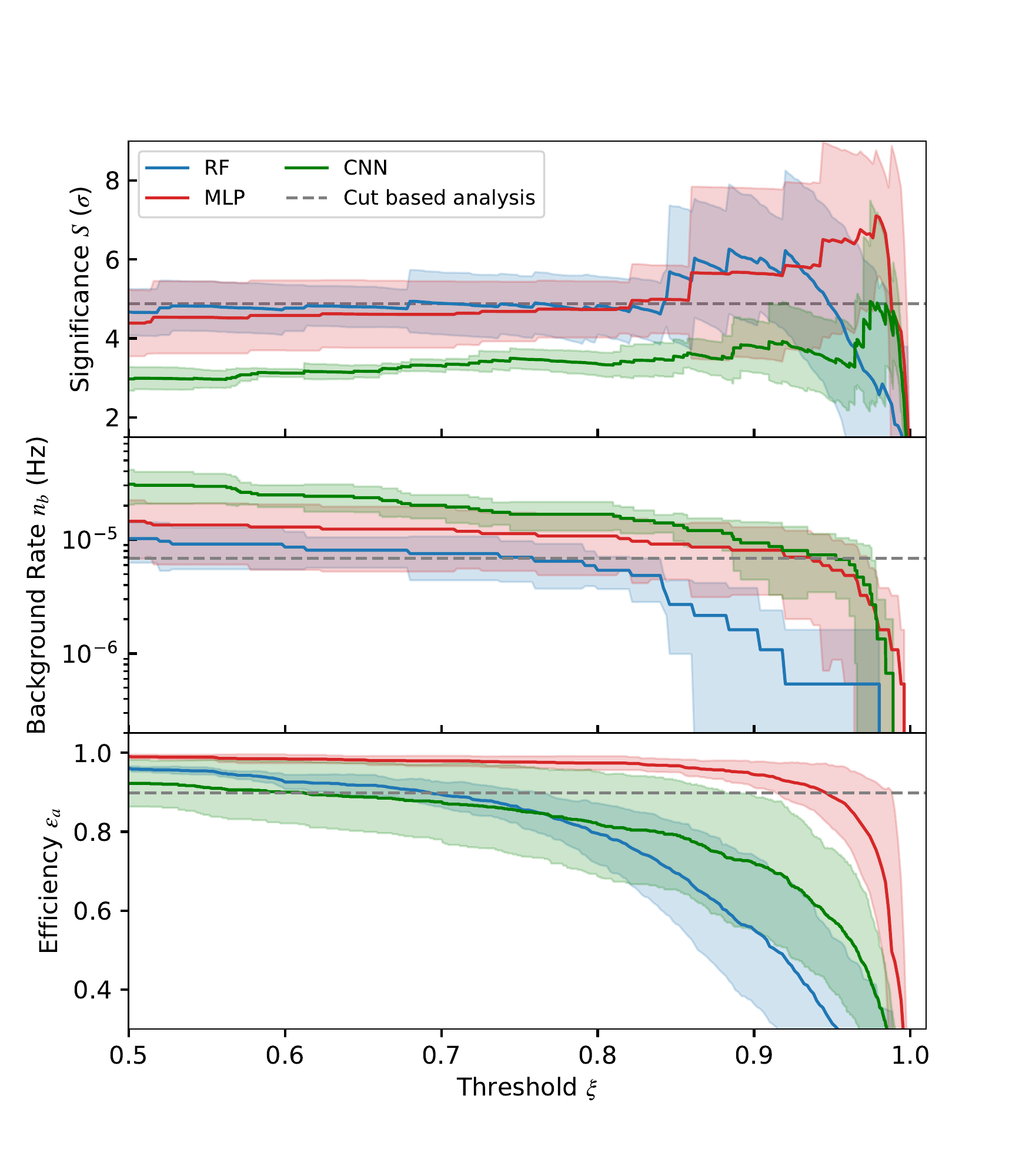}
    \caption{Performance of different classifiers (RF, MLP, and CNN) as a function of classification threshold $\xi$. Events with a predicted class label $\hat{y}_i$ will be classified as signal events if $\hat{y}_i \geqslant \xi$. The performance is shown in terms of detection significance $S$ (top), the background rate (center), and the analysis efficiency $\epsilon_a$ (bottom). The solid lines indicate the median of the performance on five different training-test splits of the data, the shaded region represent the standard deviation. The results from the cut-based analysis are shown as a dashed line.}
    \label{fig:result}
\end{figure}

The median performance of all tested classifiers on the test sets in terms of signficance $S$ (see Eq.~\eqref{eq:sig}), background rate $n_b$, and analysis efficiency $\epsilon_a$
as a function of threshold $\xi$ is shown in Fig.~\ref{fig:result}.
The shaded regions denote the standard deviation from the five different optimization runs with different test data sets. 
As expected, as $\xi$ increases, the false positive rate and thus $n_b$ is decreased as we only classify events as light-like that have predicted class labels closer to one. 
At the same time, the number of true positives and hence $\epsilon_a$ decreases as well. 
Our metric $S$ gives more weight to the false positives and as a result $S$ can be $\sim 5\,\sigma$ even for comparatively low values of $\epsilon_a$.
This can be observed in Fig.~\ref{fig:result} as well: $S$ increases with increasing $\xi$ up until the decreasing background cannot compensate the loss of true positives any longer. 
Example values for the performance are provided in Tab.~\ref{tab:result} for values $\xi$ close to maximum performance.

\begin{table}[htb]
\centering
 \caption{Classifier performance for example values of $\xi$. Values are chosen that lead to $S > 6\,\sigma$ for the RF and MLP with maximum $\epsilon_a$, whereas for the CNN the $\xi$ value is chosen that maximizes $S$. For the values of $S$, an observation time of 518\,hours and a signal rate of $2.8\times10^{-5}$\,Hz are assumed. }
 \label{tab:result}
  \begin{tabular}[htbp]{lcccc}
    \hline
    Classifier & Threshold $\xi$ & Signal efficiency & Background Rate ($10^{-6}\,$Hz) & Detection significance ($\sigma$) \\
    \hline
    Cut-based analysis~\cite{2022epsc.confE.801S} & -- & 0.898 & 6.9 & 4.88 \\
    RF & 0.862 & 0.66 $\pm$ 0.15 & 2.16 $\pm$ 2.02 & 6.04 $\pm$ 1.50 \\
    MLP & 0.944 & 0.90 $\pm$ 0.07 & 5.93 $\pm$ 5.23 & 6.51 $\pm$ 2.47 \\
    CNN & 0.974 & 0.42 $\pm$ 0.18 & $< 8.54$ & 4.94 $\pm$ 2.56\\
    \hline
  \end{tabular}
\end{table}

Our feature-based classification scheme can be compared to the performance of the cut-based analysis, which meets the ALPS~II design requirements~\cite{2022epsc.confE.801S}. 
In that analysis, the
histograms of the best-fit parameters of signal events were fit with Gaussian distributions. Using these distributions, cuts in units of Gaussian standard deviations were defined and background events were classified as such if their best-fit parameters fell outside these cut values.
It should be noted that our classifiers here provide real numbers for the class prediction, so it is in principle possible to tune $\xi$ on the training set to maximize $S$. 
The cut-based analysis presented in Ref.~\cite{2022epsc.confE.801S} did not perform a split of the data into a training and test set but reported results on the entire data set. 
Even so, our RF and MLP outperform the cut-based analysis reaching a detection significance of $\gtrsim 6\,\sigma$, albeit with large uncertainties due to the limited statistics of our data set. 
Comparing the RF and the MLP, it can be seen that the RF performs best in rejecting backgrounds whereas the MLP retains a high analysis efficiency even for high values of~$\xi$. 

In comparison to the feature-based classifiers, our CNN performs worse.
Only for high values of $\xi \gtrsim 0.97$ are we able to reach a median significance close to $5\,\sigma$ at the cost of a poor analyis efficiency with a true positive rate below 50\,\%.
The CNN performs worst of all classifiers in rejecting backgrounds and only achieves a higher true positive rate in comparison with the RF for $\xi \gtrsim 0.8$. 
It should be noted, however, that for the CNN no systematic tuning of the hyper parameters was performed and no prior knowledge of the pulse shape is required. 

\begin{figure}[htb]
    \centering
    \includegraphics[width=.5\linewidth]{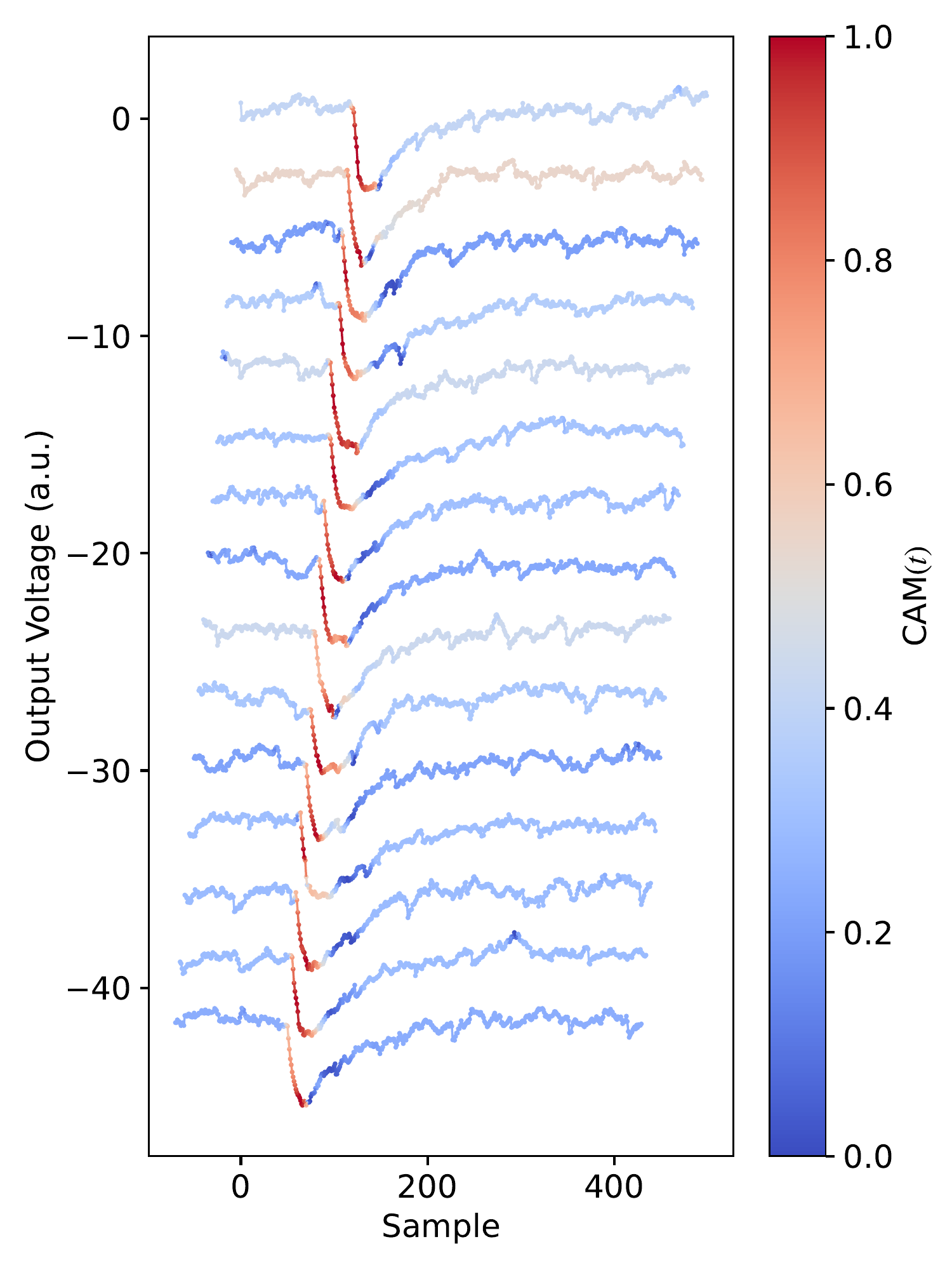}
    \caption{Class activation maps for 15 example time lines of light events which are classified as such by our CNN. The rising part of the pulse is most important for the classification of these samples. The time lines are shifted along the $y$ axis for better visibility.}
    \label{fig:cam}
\end{figure}

Figure~\ref{fig:cam} shows the CAMs defined in Eq.~\ref{eq:cam} for 15 example light pulses that were correctly classified by the network.
Higher CAM values indicate that the corresponding points are more important for classification. 
It is clearly visible that the rising part of the pulse is most important in this sense, whereas the decaying part of the pulse is less important. 
This is somewhat surprising as the background pulses in Fig.~\ref{fig:ex-event} show much longer decay times as the signal pulses.
This could be related to our choice of the kernel size: a kernel size of 11 corresponds to a time window $11 / (f_\mathrm{sample} / 4) \approx 0.9\,\mu$s and thus it is difficult for the network to capture these long trends in time. 
This might indicate an option to improve the CNN performance in the future.

\section{Discussion and Outlook}
\label{sec:conclusions}

With the low expected rate of photons reconverted from ALPs of the order of 1\,photon per day, it is of utmost importance to achieve an efficient background suppression. 
For this purpose, we have trained ML and DL classifiers on time lines measured with the ALPS~II TES detector. 
Data from a calibration setup of the TES have been used for this purpose which comprise around 1,000 real light pulses generated with a 1064\,nm laser 
and roughly 40,000 background events collected while the TES was disconnected from the optical fiber (so-called \textit{intrinsic} backgrounds). 
All our classifiers 
provide a signal-and-background discrimination that result in a potential detection significance that is higher or comparable to 
a cut-based analysis presented in Ref.~\cite{2022epsc.confE.801S}.
In particular the classifiers based on extracted features (best-fit parameters of a parametric function describing the pulse shape) can achieve 
a detection signficance in excess of $6\,\sigma$ compared to roughly $5\,\sigma$ for the cut-based analysis. 

These results are very encouraging. The present work merely serves as a proof-of-concepts and several improvements are foreseen in the future.
First, the given data set is highly imbalanced with a ratio $\sim40:1$ of background versus light data, which represents a challenge for the classifiers. 
More training data with an updated experimental setup will mitigate this problem. 
A larger set of available data will also reduce errors on the performance metrics as values of $K > 5$ for $K$-fold cross validation can be chosen while retaining large enough data sets for each iteration. 
In our tests, a CNN trained on the raw time lines performed worst. The likely reason is that a) we did not optimize the hyper parameters (e.g., number of convolutions, size of convolution kernels)
and b) the CNN might suffer most from an imbalanced data set, high frequency electronic noise, and might depend on the length of the input time lines. 
The CAMs indicate that the rising edge of the pulse is most important for discriminating signal and background events. 
The rise time could be shortened further with a higher gain bandwidth product (GBWP) of the SQUIDs. 
However, a higher GBWP will also amplify the high frequency noise. 
The reasons for this noise are currently under investigation. 

We plan to extend the present analysis on more data, in particular including background data while the optical fiber is connected to the TES, in order
to evaluate the performance of our classifiers to reject events induced by black body radiation. 
Furthermore, we will perform an optimization of the hyper parameters of the CNN and will investigate the performance of autoencoders for signal and background discrimination as done in Ref.~\cite{2019EPJC...79..450H}. 
We also plan to investigate unsupervised ML techniques in order to identify different background sources. For example, Fig.~\ref{fig:pca} suggests at least two background populations.
Lastly, it will also be interesting to see how well deep neural networks perform in reconstructing different incident photon energies and whether this can improve the energy resolution of TES detectors. 


\medskip
\textbf{Acknowledgements} \par 
M.~M.\ acknowledges the support from the Deutsche Forschungsgemeinschaft (DFG, German Research Foundation) under Germany’s Excellence Strategy – EXC 2121 ``Quantum Universe'' – 390833306 and from the European Research Council (ERC) under the European Union’s Horizon 2020 research and innovation program Grant agreement No. 948689 (AxionDM).

\medskip

%

\bibliographystyle{MSP}
\bibliography{main}





%

\end{document}